\documentclass{iau307}
\usepackage{graphicx}
\usepackage{natbib}
\usepackage{url}
\usepackage{dtklogos}
\bibpunct{(}{)}{;}{a}{}{,}

\title[Is $\lambda$ Cep a pulsating star?] 
{Is $\lambda$ Cep a pulsating star?}

\author[J. M. Uuh-Sonda et al.]   
{J.M. Uuh-Sonda$^1$, P. Eenens$^1$
 \and G. Rauw$^2$}

\affiliation{$^1$Departamento de Astronom\'ia, Universidad de Guanajuato, \\Apdo. Postal 144, 36000, Guanajuato, Guanajuato,
Mexico \\ email: {\tt juuh@astro.ugto.mx, eenens@ugto.mx} \\[\affilskip]
$^2$Institut d'Astrophysique et de G\'eophysique, Universit\'e de Li\`ege, \\All\'ee du 6 Ao\^ut, 400 Li\`ege,
Belgium \\ email: {\tt rauw@astro.ulg.ac.be} }

\pubyear{2014}
\volume{307} 
\pagerange{}
\jname{New windows on massive stars: asteroseismology, interferometry, and spectropolarimetry}
\editors{G. Meynet, C. Georgy, J.H. Groh \& Ph. Stee, eds.}

\begin{document}

\maketitle

\begin{abstract}
It has been proposed that the variability seen in absorption lines of the O6Ief star \mbox{$\lambda$ Cep} is periodical and due to non-radial pulsations (NRP). 
We have obtained new spectra during six campaigns lasting between five and nine nights. In some datasets we find recurrent spectral variations which move redward in the 
absorption line profile, consistent with perturbations on the stellar surface of a rotating star. However the periods found are not stable between datasets,  at odds with the 
NRP hypothesis. Moreover, even when no redward trend is found in a full dataset of an observing campaign, it can be present in a subset, suggesting that the phenomenon is short-lived, 
of the order of a few days, and possibly linked to transient magnetic loops.
\end{abstract}

\firstsection 
\section{Introduction}

In O-type and O-supergiant stars, line profile variability (LPV) is a widespread phenomenon \citep{Fullerton1996}. 
However, the origin of such variability is still uncertain and debatable. 
It is usually attributed to pulsations, rotational modulations, magnetic fields, structures in the stellar wind, or a mix of these phenomena.  
The presence of non-radial pulsations (NRP) has been reported mainly in late-O stars (e.g. 
HD 93521, \citealt{Rauw2008}),
while in the case of early-O stars the determination of genuine NRP has been more complicated, since their spectra are dominated 
by the so-called "red-noise" (e.g. HD 46223, HD 46150 and HD 46966, \citealt{Blomme2011}).

In 1999, de Jong et al reported the presence of low-order NRP with $P_1$= 12.3 hrs ($\ell$ = 3) and $P_2$ = 6.6 hrs ($\ell$ = 5) in the O6 Ief star, $\lambda$ Cep 
(HD 210839).
This study was conducted by analyzing the He I $\lambda 4713$ line observed for five nights in a multi-site campaign.
Since other Oef stars have displayed a strong epoch-dependence in their variability (e.g. \citealt{Rauw2003}),
the results of \citet{deJong1999} needed confirmation.

\firstsection
\section{Observations, analysis and results of our data}
We have collected 495 spectra during six observing campaigns,  four in the Observatoire de Haute Provence (OHP, France) and two in the National Astronomical Observatory of San Pedro Martir 
(SPM, Mexico). 
To analyze our datasets, we make use of the tools used by \citet{Rauw2008}. 
The results of the variability and Fourier analysis, for our He I $\lambda 4471$ and HeII $\lambda 4542$ absorption lines, are described in \citet{Uuh-Sonda2014}. 

\firstsection
\section{NRP? Long-term instability of variations}
NRP are expected to produce a pattern of alternating absorption excesses and deficits that
cross the line profile from blue to red as the star rotates \citep{Telting1997}.  
This results in a progressive variation of the phase of
the modulation across the line profile.
In our data, only a subset 
of the frequencies identified yields
a monotonic progression of the phase across those parts of the line profile where the amplitude of the variation is large. 
However, such frequencies behave differently for different epochs or lines, 
i.e. there is no long-term stability in the frequencies found. 
This suggests that we might not be observing the same mode at different epochs. Also we could not find the frequencies reported by \citet{deJong1999}. 
It appears thus that there is no evidence for persistent NRP in $\lambda$ Cep \citep{Uuh-Sonda2014}. 

\firstsection
\section{Co-rotating magnetic bright spots: short-term variability}
Among the possible causes of LPV, the idea  of ``co-rotating magnetic bright spots'' at the stellar surface 
has been gaining strength. 
Such magnetic spots would be created by magnetic fields generated in a sub-surface convective layer of massive stars \citep{Cantiello2009}, 
and their lifetime would be only of a few hours \citep{Cantiello2011}. It has also been proposed that the co-rotating magnetic bright spots 
are at the base of the so-called ``stellar prominences'', which cause cyclical variation in the wind of massive stars (e.g. for $\lambda$ Cep, \citealt{Henrichs2013}). 
As these magnetic bright spots are rotating across the line of sight, they would mimic the line profile variations usually attributed to NRP, but on short scales.
Recently, \citet{Ramiaramanantsoa2014} reported the first convincing case of co-rotating bright spots on an O-star, $\xi$ Per (O7.5III(n)((f))). 


Our analysis reveals that, in some of our campaigns, there is a persistent variability moving redward in the absorption line profile, 
consistent with perturbations on the stellar surface of a rotating star. However, in other campaigns such variability is less noticeable. 
Even during the same campaign the variability can  behave differently from day to day. 
This suggests that there is a genuine variability but it manifests itself only on short timescales, i.e. the phenomenon that causes such variability is transient.

If the variability observed in $\lambda$ Cep is produced by short-lived bright spots, some patterns could be detected in shorter datasets better 
than in longer ones. In order to test this hypothesis, we have analyzed sub-sets of our campaigns. 
This is indeed the case for example for our OHP-June-2010 He II $\lambda 4542$ line. A frequency of 0.87 d$^{-1}$ is found in the analysis of both the complete seven-day dataset and of a subset of only the first six nights. 
However only the latter shows the related redward behavior across the line profile.
This suggests a transient phenomenon that faded away during the seventh night, just as bright spots would.

We are inclined to the idea that this transitional phenomenon is actually the presence of numerous magnetic bright spots rotating with the stellar surface \citep{Uuh-Sonda2014}. 
Indeed, this idea has also been suggested by \citet{Ramiaramanantsoa2014}. However, this hypothesis still needs to be confirmed.

\firstsection
\bibliographystyle{iau307}
\bibliography{MyBiblio}

\end{document}